# A Bayesian Mean-Value Approach with a Self-Consistently Determined Prior Distribution for the Ranking of College Football Teams

James R. Ashburn[*]     Paul M. Colvert[*]

We introduce a Bayesian mean-value approach for ranking all college football teams using only win-loss data.  This approach is unique in that the prior distribution necessary to handle undefeated and winless teams is calculated self-consistently.  Furthermore, we will show statistics supporting the validity of the prior distribution.  Finally, a brief comparison with other football rankings will be presented.

[*]Atomic Football, GP, 1015 Harrison Avenue, Huntsville, AL  35801.

# 1. INTRODUCTION

In 1998, Division I-A college football took a major step forward with the introduction of the Bowl Championship Series. For the first time in history, the top two ranked teams would play for the title *by design*. For many fans, there were hopes that the seeds of a playoff system would eventually eliminate the need for ranking of teams. Unfortunately, the question of which teams would participate in a playoff still remained. The BCS committee chose to place that burden upon a composite ranking employing both voting polls and computer ranking systems. While the inclusion of computer ranking systems represented a significant stride towards unbiased rankings (at least towards individual teams, conferences, or regions), other potential biases still remained, among the more prominent being the relative importance given to margin of victory.

In 2002, the BCS committee moved to constrain the computer ranking systems to ignore margin of victory and consider only wins and losses. From a mathematical point of view, this decision was very significant. Each game was now reduced to a single bit of information. If we consider about 700 Division I-A, I-AA, II, III, and NAIA teams scheduled to play about 3600 games this year (2005), we will have 3600 bits of data at the conclusion of the season.[1] This is the equivalent of 450 bytes of information, about the same amount as one verse of "The Star-Spangled Banner." From this relatively small amount of information, we hope to accurately rank hundreds of teams. There is little wonder why the ranking of college football teams remains such a controversial problem.

Aside from the problem of limited information, another side effect of using only win/loss information is perhaps more significant. The problem arises from the fact that many ranking systems, particularly those based upon more statistical approaches, rely on some form of what is sometimes called a win probability function (WPF).[2] The win probability function describes the probability of one team winning over another given the relative *ratings* of the two teams. Note that the rating variables to which we refer are not the integer *rankings*, but instead some measure of mean *performance* calculated such that the win probability functions best match the actual game outcomes. A "best match" is often defined as the set of rating values that maximizes a composite likelihood function combining the individual WPFs. David (1988) points out that many of the more popular forms for the WPF can be written (or at least transformed to be written) as a cumulative distribution function (CDF) of the *difference* between the ratings of the two teams in a game. As a CDF, it asymptotically approaches zero on one end and unity on the other. Herein lies a problem – simply maximizing the likelihood against a set of data that includes undefeated teams will result in those teams having ratings approaching infinity. This problem has its intuitive equivalent – how does one rate undefeated teams relative to teams with one or two losses, particularly when the undefeated team plays a relatively weak schedule? After all, an upper bound to the undefeated team's performance has not really been determined. While one can assume that the undefeated team is better than all of its opponents, the question of how much better still remains.

---

[1] The limited amount of data supports the argument that rankings, for the purpose of rewarding teams with postseason play and the like, should equally weight all games. Giving greater weight to games later in the season for the purpose of determining those teams that are "best" upon conclusion of the season (similar to methods employed in so-called "predictive" models) reduces the *effective* amount of data employed in the computations.
[2] Massey (2001) uses the term "game likelihood function."

While many methods have been formulated to address the problem of undefeated and winless teams, these methods usually mark the point where the statistical foundations of many ranking systems tend to weaken. Perhaps the most statistically sound path, and one recognized by a number of authors of ranking systems on the web, is the introduction of a Bayesian prior distribution. Mease (2003) recently offered an excellent discussion of this idea. Unfortunately, a new problem arises – how does one best determine the prior distribution? Mease proposed that all teams be considered to have $n$ wins and $n$ losses each against a common virtual opponent. He states that $n$ was determined to best be unity "after testing the model on various football seasons" but no details are provided. Presumably, results were evaluated by comparisons with other ranking models. Whether or not this is the case here, comparing one model to others is a popular benchmark for validating a model, which brings us to our next topic.

## 2. THE VALIDITY OF A RANKING MODEL

A brief survey of the literature[3] seems to suggest that the primary historical method of validating ranking systems[4] is by comparison to either other ranking systems or the voting polls. While this method may be useful for identifying very poor approaches, any system being evaluated in this manner can only show itself to not be substantially worse than its peers. Kenneth Massey (whose rankings have been part of the BCS since 1999) has a very popular page on his website entitled "College Football Ranking Comparison" where he ranks ranking systems based upon the degree to which they correlate with a "consensus ranking" determined by averaging all of the systems included on the page.[5] According to Massey's website, this page is a key source for the BCS committee in their selection of computer ranking systems. To whatever degree the collective knowledge of a consensus ranking can be better than any of its individual components, such a ranking comparison has merits. Unfortunately, the fact that it is regarded as a resource for the BCS committee creates a great temptation for the authors of the ranking systems to tune their algorithms to better match the consensus. This creates a potential problem – if any substantial number of ranking systems were to be tuned for this purpose, the computer consensus becomes a "dog chasing its own tail," given that these ranking systems are also included in the consensus calculation. In effect, the consistency between the ranking systems can theoretically continue to improve completely decoupled from the degree to which the rankings actually reflect the performance of the teams. In order to elaborate on this idea, we must first introduce another metric.

---

[3] See, for example, Mease (2003), Fainmesser *et al.* (2005), Colley (2003), and Boginski *et al.* (2004).
[4] While we may, at times, use the terms ranking *system* and ranking *model* interchangeably, our preference is to use the term *model* when referring to a *system* that indeed has some underlying *model* from which it is derived. Massey (2001) appears to characterize *systems* without an underlying *model* as "formula-based systems."

This second metric, *ranking violations*, is also included on Massey's comparison page, although the ranking systems are not explicitly sorted by it. A ranking violation occurs when a team is ranked higher than another team to which it lost. This metric is often referred to in a more general sense as *retrodiction*, implying the degree to which rankings are consistent with *past* game results. At first glance, retrodiction would seem to be an excellent measure of the validity of a ranking system. However, its utility is limited for a couple of reasons. First, approaches that seek only to minimize ranking violations are unlikely to yield unique solutions.[6] Thus, an undefeated team is presumed to be ranked equally well at any point above its best opponent. Second, in the course of ranking teams by simply minimizing ranking violations, counterintuitive situations may arise. Consider the following example:

> Team A 9-1 (9 wins over team C, 1 loss to team B)
> Team B 1-1 (1 win over team A, 1 loss to team C)
> Team C 1-9 (1 win over team B, 9 losses to team A)

While a case where three teams have played 10, 2, and 10 games, respectively, might not occur in practice, the point here is to create two teams (A and C) where one team is unquestionably superior to the other. We then introduce a third team (B) that defeats the better team (A) and loses to the lesser team C. Intuitively, we might tend to rank team B between teams A and C. This would translate to attempting to balance a ranking violation in one direction with a ranking violation in the opposite direction. However, we can reduce the number of ranking violations from *two* (in the case of an ABC ordering) to *one* by ranking team B as either first (BAC) or last (ACB). In short, a minimum ranking violation approach would dictate that team B is either best or worst but not in between. This is obviously a paradoxical conclusion.

---

[5] See http://www.masseyratings.com/compare.htm.
[6] According to Jay Coleman, "there are literally trillions of different rankings at any given point in time that would yield the same minimum number of violations" (see http://www.unf.edu/~jcoleman /minv.htm).

Does this paradox occur in practice? One result that is often seen in computer rankings but seen with markedly less frequency in polls is where a team is ranked *slightly* below a team it defeated. When such a situation occurs, fans will often cry foul, invoking the concept of "head-to-head" competition. Similarly, voters in the polls respond to naturally "correct" such situations, flipping the closely ranked teams to alleviate the perceived inconsistency. The typical human response, however, is flawed. These ranking violation "corrections" are most often demanded when the two teams involved fall closely in the polls. Curiously though, ranking violations involving teams sufficiently separated in the polls seem to be comfortably ignored. In effect, big upsets are more easily believed than small ones.[7] Obviously, were two teams involved in a ranking violation to fall consecutively in the rankings, why *not* "correct" the violation? After all, a correction in this situation cannot induce other ranking violations. However, the implications of this line of reasoning are counterintuitive, suggesting that if team A defeats team B, then team A is either much inferior to team B (as determined by other games played by both teams) or team A is at least slightly superior to team B (as determined by the "head-to-head" match), but team A cannot be slightly worse than team B. As we will soon show, upsets are not only a natural part of college football, they are extremely common, occurring about 20% of the time. Thus, ranking violations should be expected (including these "minor violations"), and it should not simply be the purpose of a ranking system to minimize them but to instead be able to estimate the frequency of them.

Now that we have introduced retrodiction, we will return to our discussion of the potential problem mentioned above – the risk of ranking systems being tuned to match a consensus of which they are a part, resulting in the computer consensus "chasing itself." While we cannot verify a *cause*, there is perhaps evidence for the *effect*. Initial analysis performed by us indicates that Massey's computer consensus may indeed be undergoing a transition of this kind. Among the indicators is an apparent shift in the correlation between the "correlation to consensus" scores and the "ranking violation" scores. Assuming the majority of these algorithms are not based upon explicitly minimizing ranking violations, then one would expect that the "better" systems, as measured by a high correlation with the consensus, would tend to have a lower occurrence of ranking violations. Presumably, this general relationship should be detectable when a sufficient number of ranking systems are examined. In other words, there should be a *negative* correlation between the "correlation to consensus" scores and the respective "ranking violations." However, Figure 1 shows that from 2000 to 2005, there appears to be a trend (as indicated by the arrow) in which the highest ranked ranking systems show an increasingly *positive* correlation. In 2000, a consistent negative correlation is detectable when about 20 or more of the highest ranked systems are examined. By 2005, however, a consistent positive correlation becomes apparent when the 35 highest ranked systems are considered, and the coefficient remains positive until about 70 systems are included. This suggests the possibility that the consensus is beginning to decouple with actual outcomes.

---

[7] Of course, we define an upset here based upon a team's mean performance/rating for a season. There is little doubt that some teams strengthen during the course of a season while others wane. It is currently beyond the scope of this paper to attempt to quantify this effect. Furthermore, attempting to do so within any ranking model will at least double the number of unknowns that must be estimated from the same data, although this may admittedly be of value to predictive models.

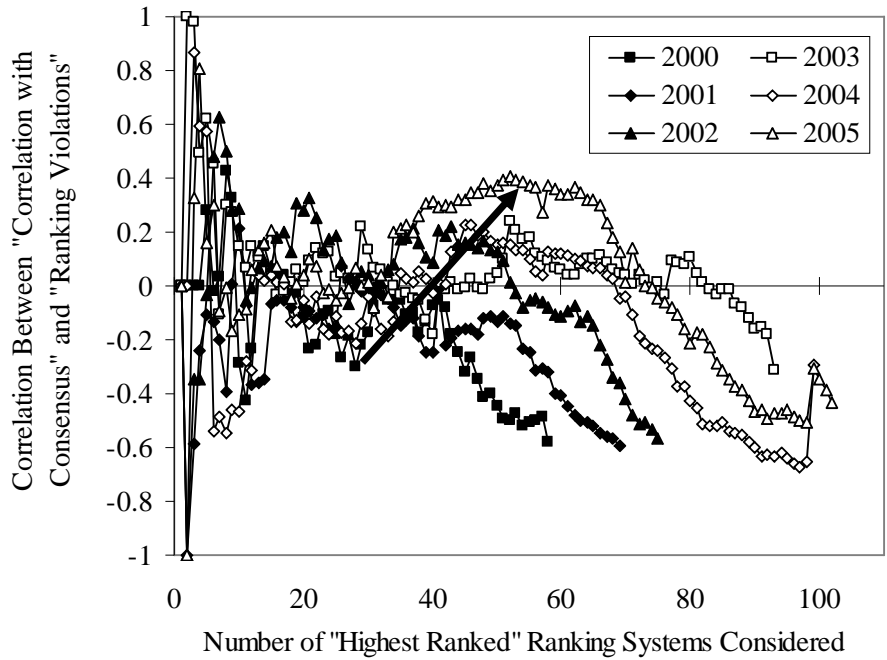

**Figure 1.** Correlation ($\rho$) Between the "Correlation to Consensus" Score and the "Ranking Violations" Score for the N "Highest Ranked" Ranking Systems

Thus, we suggest that the more popular metrics for validating ranking systems have potential weaknesses, and we are still faced with the dilemma of how to unambiguously validate a ranking system. While we do not claim to offer a complete solution, we would like to suggest a different path. In doing so, we begin by making the following proposition:

> *A valid ranking system should, first of all, be based upon a model. As such, the algorithm derived from it should be able to determine metrics that can be validated against equivalent metrics that can be independently derived from the game data.*

We will attempt to clarify this and concurrently achieve it in the analysis section below. However, we must first introduce the model.

## 3. THE MODEL

The model introduced here closely parallels the model documented by Mease (2003). Other models with similar features -- Bayesian prior, WPF based upon a normal CDF per Thurstone (1927) and Mosteller (1951), etc. -- can be found documented on the web by Massey (2001) and Dolphin,[8] among others. *The fundamental difference between the model herein and other similar models is the method by which the prior distribution is obtained.* For this reason, we will move quickly through an introduction of the basics of the model (referring the reader to Mease's work for more details) in order to focus our attention on the derivation of the prior distribution.

Assume that over the course of a season, each team $i$ can be characterized by a single "mean performance" variable $r_i$. From game to game, each team's "effective performance" $r_i'$ is assumed to deviate about the mean in a manner described by a normal distribution of variance 1/2. We shall call this variance the *performance variability*. The specific value chosen is irrelevant (defining only the scale of the final ratings) and has been selected simply for convenience.[9] Assuming the performance deviations of two teams in a game are independent,[10] the *relative performance variability* associated with the relative performance $r_i' - r_j'$ becomes

---

[8] See http://www.dolphinsim.com/ratings/.
[9] Mease (2003) independently selected the same value.
[10] To the degree that teams "play up" or "play down" to their competition can, in the mean, be modeled by a simple scale factor between the mean performance values of teams in a game, such an effect simply rescales the ratings results, with no effect on the final rankings. Given that the scale is arbitrary, only second order effects would then affect the final rankings. Such a second order effect would be, for example, that teams are more prone to "play up" to competition of comparable

$$\sigma_v = \sqrt{\frac{1}{2} + \frac{1}{2}} = 1, \tag{1}$$

thus revealing the rationale for the choice of scale.

In a game between team $i$ and team $j$, team $i$ is assumed to win when $r_i' > r_j'$ (or $r_i' - r_j' > 0$).[11] Thus, the probability of team $i$ defeating team $j$ is given by the win probability function

$$P(r_i - r_j, \sigma_v) \tag{2}$$

where $P(x, \sigma)$ is the cumulative normal distribution[12]

$$P(x, \sigma) = \int_{-\infty}^{x} p(x, \sigma) dx \tag{3}$$

and $p(x, \sigma)$ is the normal distribution

$$p(x, \sigma) = \frac{1}{\sqrt{2\pi}\sigma} \exp\left(-\frac{x^2}{2\sigma^2}\right). \tag{4}$$

We will now formally define the term *rating* to refer to the mean performance of a team, whether true or estimated. Let us also define the following:

    $i$ = team index
    $r_i$ = true rating of team $i$
    $\tilde{r}_i$ = estimated rating of team $i$
    $\sigma_i^2$ = estimated uncertainty of team $i$'s rating
    $N_i$ = number of games scheduled by team $i$
    $j$ = index indicating teams *played* by team $i$

---

performance – when the likelihood of the added effort affecting the game outcome is highest – and less likely to do so when the performance differential is high.

[11] Neglecting home field advantage, which we will address in a later section.

[12] While Mease (2003) argues for the validity of the normal CDF (based upon the central limit theorem), others contend that the logit CDF -- functionally equivalent to the Zermelo model, aka the Bradley-Terry (1952) model -- is best suited for most real world "binary symmetric games." See, for example, Smith (1994). In considering Smith's arguments, note that he did not point out that the logit CDF can be derived from a model where teams have *independent exponentially* distributed performance distributions.

$k =$ index indicating teams *scheduled* by team *i* (including teams played)

Two subscripts, *ij* for example, indicate a reference to a game between teams *i* and *j*. In some contexts, *ij* and *ik* may be abbreviated as *j* and *k*, respectively, where *i* is understood. The operation $\langle \ \rangle_k$ indicates a uniformly weighted average over all scheduled games.

Given these definitions and assumptions, let us assume that we have uncertain estimates of the ratings for team *i*'s opponents. In order to estimate team *i*'s rating, we will need to develop the WPF as part of an integrand with which we will integrate over all possible values of team *i*'s rating:

$$\Phi_{ij}(r) = \begin{cases} P(\Delta_j(r), \sigma_{jv}) & \text{if } i \text{ wins} \\ P(-\Delta_j(r), \sigma_{jv}) & \text{if } j \text{ wins} \\ p(\Delta_j(r), \sigma_{jv}) & \text{if a tie} \end{cases} \qquad (5)$$

where

$$\Delta_j(r) = r - \tilde{r}_j \qquad (6)$$

and

$$\sigma_{jv} = \sqrt{\sigma_j^2 + \sigma_v^2} = \sqrt{\sigma_j^2 + 1}. \qquad (7)$$

Note that the first two forms for $\Phi_{ij}(r)$ are cumulative distribution functions while the third is the normal distribution function. While the latter should be unnecessary for modern college football with the universal use of overtime, it has been included here for completeness.

We now have a form for the WPF $\Phi_{ij}(r)$ where it will be used to *iteratively* estimate the rating of team *i* given *estimates* of the ratings $\tilde{r}_j$ of its opponents, thus the inclusion of the opponent's rating uncertainty. Equation (7) represents the combined effects of the relative performance variability from equation (1) and the uncertainty of the opponent's rating $\sigma_j^2$ to be derived below.

Given these definitions, we define a set of intermediate moments

$$m_i(n) = \int_{-\infty}^{+\infty} dr \, r^n \Phi_p(r, \sigma_p) \prod_j \Phi_{ij}(r) \qquad (8)$$

where n = 0, 1, 2, and where we have introduced the very important prior distribution $\Phi_p(r,\sigma_p)$ discussed earlier. $\Phi_p(r,\sigma_p)$ will be derived in the next section.

With these definitions, we can now estimate each team's rating by the following:

$$\tilde{r}_i = \frac{m_i(1)}{m_i(0)} \qquad (9)$$

where the zeroth moment is necessary for proper normalization of the result. Furthermore, the uncertainty in the rating can be estimated by

$$\sigma_i^2 = \frac{m_i(2)}{m_i(0)} - \left(\frac{m_i(1)}{m_i(0)}\right)^2. \qquad (10)$$

Thus, once we have an equation for the width of the prior distribution, we will have a complete set of self-contained, self-consistent equations that can be solved iteratively for the team ratings. It is important to note here that maximum likelihood approaches are far more popular for football rankings than those based upon mean values. While maximum likelihood solutions often reduce or eliminate the need for numerical integration, they also impair the potential utility of the higher order statistics such as rating uncertainty estimates. As will become apparent in the derivations that follow, a mean value approach and the options it affords are critical to our model.

Before describing the process for determining the prior distribution, we will take a moment to further discuss its necessity. Consider equation (8) less the prior distribution. If a team *i* is undefeated or winless, the composite distribution

$$\prod_j \Phi_{ij}(r) \qquad (11)$$

will have but a single tail, resulting in moments (and ratings) of infinite magnitude. Thus, without a confining prior distribution, all undefeated teams will be estimated to be of infinite performance. Based strictly upon a model employing only wins and losses, the mean (and maximum likelihood) performance must be infinite, leaving all undefeated teams indistinguishable. Similarly, winless teams will be estimated to be of infinite negative rating. It is apparent that such a problem is most serious in sports where conditions are conducive to undefeated and winless teams, namely relatively small numbers of games and often uncompetitive schedules. College football is perhaps most prominent in this regard, while sports such as professional baseball might fair quite well without the prior distribution.

A survey of the internet has revealed a few rating methods that are similar to ours up to this point in the derivation. A handful of which we are aware have introduced prior distributions, but, to our knowledge, none of the parameters in the prior distributions described in those models appear to have been derived in any self-consistent manner. Furthermore, we are not aware of any where attempts have been made to quantitatively demonstrate the validity of the prior distributions. Thus, it is here where we suggest that our model is a significant step towards achieving a truly self-contained model of its kind.

## 4. THE PRIOR DISTRIBUTION

We begin this section with the following conjecture:[13]

> *In the presence of large numbers of games per team, a prior distribution can be easily estimated but is relatively unnecessary. In the presence of very small numbers of games per team, a prior distribution cannot be easily determined but is critical to achieving well-behaved results. In the presence of intermediate numbers of games per team, it may be possible to describe the relationships between the posterior and prior distributions from which a self-consistent prior distribution can be estimated.*

In order to elaborate on this supposition, we begin with an intermediate (and simpler) form for what will eventually become our prior distribution. Let us assume that the ratings across all teams are normally distributed according to

$$\Phi_p(r, \sigma_p) = \frac{1}{\sqrt{2\pi}\sigma_p} \exp\left(-\frac{r^2}{2\sigma_p^2}\right) \tag{12}$$

---

[13] AUTHOR'S NOTE: This paper in its current form was developed prior to our introduction to Empirical Bayes methods. Until we have the opportunity to replace the derivations below with a rigorous application of empirical Bayes, we provide in the interim the original text (albeit a somewhat clumsy and parochial "rediscovery" and application of empirical Bayes). The reader may note that the analysis which follows suggests that the hyperparameters (another new term to us) which we derived are likely not particularly poor estimates of those we hope to derive more precisely in the future.

where $\sigma_p$ is as yet undetermined. Now assume that no games have yet been played. Then, using equations (9) and (10) we find the following

$$\tilde{r}_i = 0 \tag{13}$$

and

$$\sigma_i^2 = \sigma_p^2 \tag{14}$$

for all teams. Next, consider the other extreme case where all the teams have played an infinite number of games. We then expect each team's estimated rating to converge to its true rating. If the variance of the prior estimate from equation (12) is correct, then the variance of the rating estimates will be

$$\left\langle \tilde{r}_i^2 \right\rangle = \sigma_p^2 \tag{15}$$

where, for the purposes of this argument, we are assuming that the mean estimated rating is defined to be exactly zero. Given an infinite number of games, the corresponding rating uncertainties will approach zero,

$$\sigma_i^2 = 0. \tag{16}$$

Thus, we hypothesize that

$$\sigma_p^2 = \left\langle \tilde{r}_i^2 \right\rangle + \left\langle \sigma_i^2 \right\rangle \tag{17}$$

in all cases, including intermediate (non-zero, finite) numbers of games. Another way to view this is that the prior distribution can be represented by the superposition of the distributions that are described by the rating estimates and their corresponding rating uncertainties. In the previous equation, we have assumed that all teams schedule and play the same number of games, an issue we will address in greater detail later.

If teams randomly selected their opposition from among all teams, then our current form for the prior distribution should be valid. However, given that most games are within the same division and conference, we will now expand our prior distribution to the following form

$$\Phi_p(r,\sigma_p) = \frac{1}{\sqrt{2\pi}\sigma_p} \exp\left[-\frac{\left(r - \langle r_{ik} \rangle_k\right)^2}{2\sigma_p^2}\right] \qquad (18)$$

We have now centered the prior distribution on the mean *scheduled* opponent's rating. While we could have centered it on the mean *played* opponent, by the end of the season the two will be equivalent anyway. Furthermore, prior to the end of the season, this form appears to lend greater stability to the midseason ratings.

To estimate $\sigma_p$ for this new form of the prior distribution, we will need to determine two other quantities. The *first* quantity will measure the variance of ratings within the typical schedule. It will be referred to as the *mean schedule variance* (MSV). A small value indicates that teams tend to schedule opponents of very similar performance to each other. A large value indicates that teams schedule opponents of widely varying performance. The *second* quantity will measure how well teams compare to their mean opponent's rating. It will be referred to as the *mean team variance* (MTV). A small value indicates that teams tend to be well matched to their "average opponent." A large value indicates that teams are often poorly matched to their "average opponent."

Before we proceed further, let us examine the significance of the relative magnitudes of the mean schedule variance and the mean team variance. First, consider the case where the MSV is much larger than the MTV. In such a case, teams will more consistently play schedules where about half of their opponents are inferior and half are superior. Thus, winning percentages will have a tendency to cluster around 0.500. Next, consider the case where the MSV is much smaller than the MTV. In this case, a significant number of teams will play schedules where all of their opponents are either substantially inferior or substantially superior. Thus, winning percentages will have a wide variation, with many teams going undefeated or winless.

Therefore, we would expect the relative magnitudes of the MSV and the MTV to be reflected in the distribution of winning percentages. In fact, the variance of winning percentages was our first attempt at a second metric for validating our estimates of the MSV and MTV. However, because the distribution of winning percentages is sensitive to the numbers of games played by the various teams (which often ranges from about nine to thirteen games in a typical season), we formulated a related but more practical metric which we call the *outcome correlation coefficient*. This metric is effectively a correlation coefficient describing the degree to which the outcomes of two games are correlated (i.e., win-win/loss-loss vs. win-loss) where one team is common to both games. Ultimately, we wish to evaluate this metric across all valid game pairs across all teams, thus yielding a quantity that characterizes a given season.

We will start by deriving the outcome correlation coefficient for a single team. Note that for any given team, the number of *independent* game pairings is effectively the number of games less one. While we could pair games according to the order in which they were played, this might make the result vulnerable to "trends" in performance over the course of a season. Since the result will obviously vary depending upon how the games are paired, we will instead evaluate the coefficient across all possible pairings within each team's schedule and then weight that team's contribution to the season's coefficient by the effective number of independent pairings. As a matter of expediency, we define a win by the team common to the pair of games as an outcome of 1.0 and a loss as an outcome of -1.0. Thus, the correlation coefficient for a team $i$ with $W_i$ wins and $L_i$ losses is

$$\rho_i = \frac{\frac{1}{2}W_i(W_i-1) + \frac{1}{2}L_i(L_i-1) - W_i L_i}{\frac{1}{2}(W_i+L_i)(W_i+L_i-1)}. \tag{19}$$

For the sake of clarity, we have neglected to simplify the result. From this equation, one can see that $\rho$ is 1.0 for undefeated and winless teams, approaches zero for 0.500 teams as the number of games goes to infinity, and is -1.0 only for teams with exactly one win and one loss. It is, of course, indeterminate for teams with fewer than two games. In order to achieve well-behaved results, we generate a weighted average outcome correlation coefficient for a season according to

$$\rho = \frac{\sum_i (W_i + L_i - 1)\rho_i}{\sum_i (W_i + L_i - 1)}$$

$$= \frac{\sum_i \frac{\frac{1}{2}W_i(W_i-1) + \frac{1}{2}L_i(L_i-1) - W_i L_i}{\frac{1}{2}(W_i+L_i)}}{\sum_i (W_i + L_i - 1)}. \tag{20}$$

We now have a parameter that measures overall *scheduling parity*, that is, how well teams are matched to their opposition. It also serves as a measure of the relative magnitudes of the mean team variance and the mean schedule variance. A value approaching unity indicates that the MTV is much larger than the MSV (see Figure 2), while a value approaching zero indicates that the MTV is much smaller than the MSV (see Figure 3).

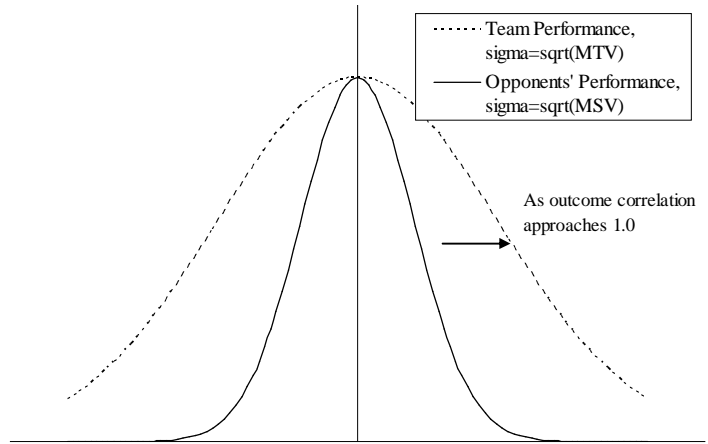

**Figure 2.** Relationship between the Outcome Correlation Coefficient and the Relative Magnitudes of Mean Schedule Variance (MSV) and Mean Team Variance (MTV) as the MTV Increases.

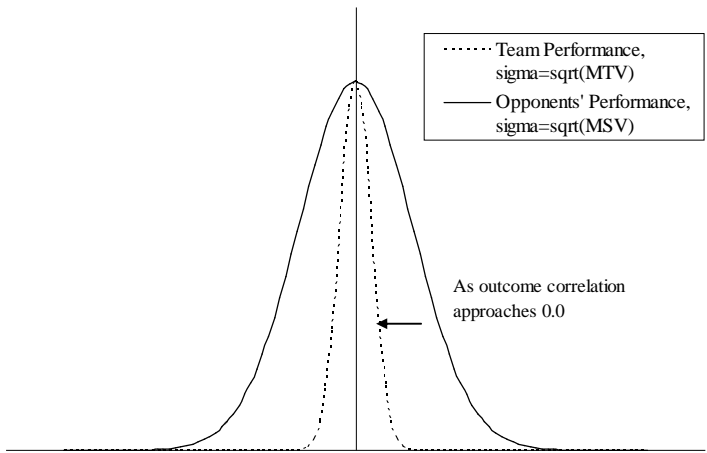

**Figure 3.** Relationship between the Outcome Correlation Coefficient and the Relative Magnitudes of Mean Schedule Variance (MSV) and Mean Team Variance (MTV) as the MTV Decreases.

Table 1 below shows the values of the outcome correlation coefficient derived for recent years.

| Season | Outcome Correlation Coefficient ($\rho$) | Estimated "Precision" ($\sigma_\rho$) |
|---|---|---|
| 2001 | 0.164 | 0.011 |
| 2002 | 0.158 | 0.011 |
| 2003 | 0.164 | 0.010 |
| 2004 | 0.151 | 0.010 |
| 2005[*] | 0.157 | 0.011 |

[*]Games through 11/19/05.

**Table 1.** Outcome Correlation Coefficient for all Division I-A, I-AA, II, III, and NAIA College Football Teams for the Years 2001-2005.

Without going into details, the quantity $\sigma_\rho$ is a coarse estimate of the "precision" of the correlation coefficient based upon the premise that a season represents a sample of some hypothetical large population and that if the given season could be replayed multiple times, with the teams maintaining their mean performance levels, then the coefficient would tend to vary according to the indicated precision. Note that the estimates of $\sigma_\rho$ are on the order of the seasonally variability in the correlation coefficient (about 0.005 in this small sample), indicating that *scheduling parity* has not varied to any statistically significant degree over this relatively brief period.

To this point, we have not been particularly concise about distinctions between sample and population variances. In our model, we presume that a season of games represents a sampling of errored "measurements" of relative performance from some potentially infinite hypothetical population. As more "measurements" (i.e., games) are taken, the accuracy with which true mean performance can be measured improves. We will also assume that schedules represent samples as well. Thus, if more games could be scheduled and played, the "true" mean opponents' ratings could be measured with increasing accuracy. Given the relatively small samples inherent in college football (and football in general), we must be careful in the following derivations to account for the differences between sample and population variance. This is also necessary to account for the effects of schedules with differing numbers of games.

We will defer until later in our discussion a derivation of the relationship between the outcome correlation coefficient and the MSV and MTV. We will proceed instead with the derivation of the MSV and MTV from within the ranking model itself, followed by a demonstration of how the prior distribution introduced in equation (18) is a function of these variables.

First, we determine the sample variance of each team's scheduled opponents' ratings about the mean opponent's rating

$$s_i'^2 = \left\langle \left( r_{ik} - \langle r_{ik} \rangle_k \right)^2 \right\rangle_k. \tag{21}$$

Note that this expression is currently in terms of the true ratings. Later, we will translate it (and several of the expressions to follow) into forms operating on the rating estimates – where we will be required to consider also the effect of the rating uncertainties.

Now, since this sample variance is a biased estimator[14] that is sensitive to the number of scheduled games played by each team, we cannot determine a meaningful average of this quantity across teams without an equivalent unbiased estimator, in this case the equivalent population variance as given by

$$\sigma_i'^2 = \frac{N_i}{N_i - 1} s_i'^2. \tag{22}$$

Because the variance of this estimator, var($\sigma_i'^2$), is inversely proportional to $N_i$-1,[15] we will generate a well-behaved least-squares weighted average – our "mean" schedule variance – as follows:

$$\sigma'^2 = \frac{\sum_i (N_i - 1)\sigma_i'^2}{\sum_i (N_i - 1)} = \frac{\sum_i N_i s_i'^2}{\sum_i (N_i - 1)}. \tag{23}$$

This pattern of $N_i$-1 weighting will be used a number of times in the derivations to follow. Note that in the last few equations, the prime symbol is used to denote the "schedule variance," that is, the distribution of ratings within a schedule about the mean rating.

---

[14] See, for example, Mathworld, http://mathworld.wolfram.com/SampleVariance.html.
[15] Assuming underlying normal distributions. See Mathworld, http://mathworld.wolfram.com/k-Statistic.html and http://mathworld.wolfram.com/NormalDistribution.html.

Similarly, we wish to determine the distribution of the ratings of the teams themselves about their mean opponents' ratings. For each team, the sample variance is but a single "measurement"

$$s_i^2 = \left(r_i - \langle r_{ik} \rangle_k\right)^2. \tag{24}$$

Just as the prime symbol above indicated schedule variance, its absence here indicates team variance. Correcting for the bias between sample and population that comes through the mean opponents' rating yields

$$\sigma_i^2 = s_i^2 - \frac{s_i'^2}{N_i - 1}. \tag{25}$$

Weighting by $N_i$-1 to yield the mean team variance produces

$$\sigma^2 = \frac{\sum_i (N_i - 1)s_i^2 - s_i'^2}{\sum_i (N_i - 1)}. \tag{26}$$

In order to replace the true ratings with the estimated ratings, we must include the effects of the rating uncertainties. Assuming that game-to-game performance deviations are independent yields the following substitutions:

$$r_i \to \tilde{r}_i \tag{27}$$
$$r_{ik} \to \tilde{r}_{ik} \tag{28}$$
$$s_i'^2 \to s_i'^2 + \frac{N_i - 1}{N_i} \langle \sigma_{ik}^2 \rangle_k \tag{29}$$
$$s_i^2 \to s_i^2 + \sigma_i^2 + \frac{\langle \sigma_{ik}^2 \rangle}{N_i}. \tag{30}$$

Given that the final terms in the latter two expressions would cancel neatly upon substitution into equation (26) above, we were very tempted to proceed with this form. However, from the form of equation (8) above, it is clear that were all of a team *i*'s opponents' ratings errored by the same quantity, then the rating for team *i* would be offset by the same amount, as both the prior distribution and the win probability functions would be similarly translated. In the simplest possible model of this correlation, [16] we simply eliminate the final term in equation (30) yielding

$$s_i^2 \rightarrow s_i^2 + \sigma_i^2. \tag{31}$$

Completing all the substitutions yields the MSV ($\sigma'^2$) and MTV ($\sigma^2$), respectively,

$$\sigma'^2 = \frac{\sum_i N_i \left[ \left\langle (\tilde{r}_{ik} - \langle \tilde{r}_{ik} \rangle_k)^2 \right\rangle + \frac{N_i - 1}{N_i} \langle \sigma_{ik}^2 \rangle_k \right]}{\sum_i (N_i - 1)} \tag{32}$$

$$\sigma^2 = \frac{\sum_i (N_i - 1)\left[ (\tilde{r}_i - \langle \tilde{r}_{ik} \rangle_k)^2 + \sigma_i^2 \right] - \left[ \left\langle (\tilde{r}_{ik} - \langle \tilde{r}_{ik} \rangle_k)^2 \right\rangle + \frac{N_i - 1}{N_i} \langle \sigma_{ik}^2 \rangle_k \right]}{\sum_i (N_i - 1)} \tag{33}$$

which are in terms only of the team ratings, rating uncertainties, and numbers of games scheduled.

Analogous to equations (22) and (25), the estimates of the *prior* team and schedule sample variances ($\sigma_p^2$ and $\sigma_p'^2$, respectively) for each team should follow from

$$\sigma'^2 = \frac{N_i}{N_i - 1} \sigma_p'^2 \tag{34}$$

and

$$\sigma^2 = \sigma_p^2 - \frac{\sigma_p'^2}{N_i - 1}. \tag{35}$$

Eliminating $\sigma_p'^2$ and solving for $\sigma_p^2$ yields

---

[16] This can be derived by assuming that the team rating error includes a component that is the average of the errors in the ratings of its opponents. This results in cross-terms that will effectively cancel out the final term in equation (30).

$$\sigma_p^2 = \sigma^2 - \frac{\sigma'^2}{N_i}, \tag{36}$$

which becomes our method of estimating of the variance in the prior distribution of equation (18). Note that in equation (18) we did not anticipate that the variance of the prior distribution would vary from team to team. From the form above, however, it is clear that it must vary between teams according to the number of scheduled games.

We have gone to great lengths to derive a prior distribution centered on a mean opponent. Given division and conference play, we believe that this is a valid model, and we will support this with the results we shall derive below. Of course, this might not necessarily be a good model for all sports. If typical schedules were representative of the full set of teams (for example, if opponents were randomly selected from all teams), then a prior distribution centered on some arbitrary point (e.g., zero) would be more valid. This was, in fact, the form we derived previously in equation (17).

In any case, we now possess a complete set of equations. The unknowns are the team ratings, the rating uncertainties, the mean schedule variance, and the mean team variance from which the variance of the prior distribution can be determined for each team. To solve the equations, we do so iteratively. All teams are given an initial rating of zero and rating uncertainty of one, and the MSV and MTV are each set to one. The team ratings and rating uncertainties are updated based upon the game outcomes, from which new estimates for the MSV and MTV are determined. The process is repeated until convergence is achieved. As a practical matter, because the ratings are not anchored to any absolute point, only the relative ratings (i.e., differences) are significant. As a result of this, there is often a slight tendency in the course of numerical simulation for the ratings to drift slightly as a group, thus frustrating attempts to determine convergence. Thus, at the conclusion of each iteration, a mean rating is determined and then subtracted from all ratings, restoring an average rating of zero.[17] This "normalized" result can then be compared to the previous iteration to determine if convergence has been achieved.

---

[17] It is also necessary that there be no disconnected subgroups. For example, the Division III NESCAC Conference teams play only games among themselves and must therefore be excluded for convergence to be achieved.

## 5. ANALYSIS

From the MSV and MTV in equations (32) and (33), respectively, we can estimate an outcome correlation coefficient for the prior distributions they represent

$$\rho = \int_{-\infty}^{+\infty} dr_i \int_{-\infty}^{+\infty} dr_j \int_{-\infty}^{+\infty} dr_k \Phi(r_i,\sigma)\Phi(r_j,\sigma')\Phi(r_k,\sigma') \times \\ [1 - 2P(r_i - r_j)][1 - 2P(r_i - r_k)] \tag{37}$$

where we have exploited the fact that

$$P(i \text{ defeats } j)P(i \text{ defeats } k) + P(j \text{ defeats } i)P(k \text{ defeats } i) \\ - P(i \text{ defeats } j)P(k \text{ defeats } i) - P(j \text{ defeats } i)P(i \text{ defeats } k) \\ = P(r_i - r_j)P(r_i - r_k) + P(r_j - r_i)P(r_k - r_i) \\ - P(r_i - r_j)P(r_k - r_i) - P(r_j - r_i)P(r_i - r_k) \\ = [P(r_i - r_j) - P(r_j - r_i)][P(r_i - r_k) - P(r_k - r_i)] \\ = [1 - 2P(r_i - r_j)][1 - 2P(r_i - r_k)]. \tag{38}$$

Applying our algorithm to the results from the 2001-2005 seasons and then estimating the outcome correlation coefficient from the final values for the MSV and MTV, we obtained the following results (where we have repeated here the previous results from above to expedite the comparison):

| Season | Actual $\rho$ | Estimated $\rho$ | Difference |
|--------|---------------|------------------|------------|
| 2001   | 0.164         | 0.161            | -0.003     |
| 2002   | 0.158         | 0.158            | 0.000      |
| 2003   | 0.164         | 0.164            | 0.000      |
| 2004   | 0.151         | 0.153            | +0.002     |
| 2005*  | 0.157         | 0.151            | -0.006     |

*Games through 11/19/05.

**Table 2.** Outcome Correlation Coefficient for all Division I-A, I-AA, II, III, and NAIA College Football Teams for the Years 2001-2005 Derived from Actual Game Data and from Prior Distributions Obtained from the Model.

From the results, it is clear that the estimates are both quite accurate and relatively unbiased.

We have not noted in the table above the values derived for the MSV and MTV. These are, however, worthy of some discussion. For the years 2001-2005, the square root of the MSV ranged from about 0.93 to 1.02 with no discernable trend. The square root of the MTV ranged from about 0.76 to 0.85. Prior to being able to calculate these quantities, we speculated as to their relative magnitude. A value of the MTV much larger than that of the MSV seemed counterintuitive – why would teams tend to often schedule opponents either all much better or all much worse than themselves? Thus, it seemed reasonable that the MTV would be no larger than the MSV. Furthermore, it seemed reasonable that teams would desire to be well-matched to their opposition.[18] For this reason, one might expect the MTV to be much smaller than the MSV. However, because conference play usually accounts for more than half of all games, the MTV could not be substantially smaller. Thus, our final expectation was that the MTV would be comparable to or possibly slightly smaller than the MSV, which we indeed found.

Unfortunately, validation of the prior distribution in our case is not complete. While the outcome correlation coefficient describes the magnitudes of the MSV and MTV relative to each other, a second metric is required to specify them uniquely. More concisely, a second metric is required to describe the magnitudes of the MSV and MTV relative to the performance variability from equation (1). Formulating a purely objective metric has proven frustrating. In the interim, we can only bound the results as follow.

Consider the case where the MSV and MTV are much larger than $\sigma_v$. In this case, games will tend to match teams with performance differences much larger than $\sigma_v$. Thus, upsets – games where the "better" team loses – will be very rare. By "better," we refer to the *true* rating. Consider now the other extreme case where the MSV and MTV are much smaller than $\sigma_v$. In this case, games will often match teams with performance differences much smaller than $\sigma_v$, and upsets will consequently be very common.

Unfortunately, unlike the outcome correlation coefficient, the frequency of upsets cannot be determined directly from the game results because we must know the ratings to determine which team in a game is "better." Even less fortunate, the estimated ratings from the model itself are insufficient for this purpose since they are both uncertain, and they represent a biased[19] posterior distribution, not a prior one.

---

[18] One might argue that teams would desire an advantage over their opponents, but obviously this cannot be achieved on average across the entire population.

One alternative metric that was explored involves sets of three teams that each play the other two. If upsets are frequent, then it would be common to find cases where each of the three teams goes 1-1. Infrequent upsets would mean that the teams most often go 2-0, 1-1, and 0-2. Unfortunately, there appears to be no means to demonstrate that teams that play these "round robins" are representative samples. In fact, we suspect that they are specifically unrepresentative of these distributions since they would often be confined to conference play where teams tend to be better matched than in nonconference play. Since we have not yet been able to formulate a second metric with the suitable properties, we must concede that we have not fully achieved what we proposed at the end of Section 2. In the meantime, we can only offer a "sanity check" of sorts on our results using the frequency of upsets as our metric.

Using the MSV ($\sigma'^2$) and MTV ($\sigma^2$), we can estimate the frequency of upsets as

$$U = \int_{-\infty}^{+\infty} dr_i \int_{-\infty}^{+\infty} dr_j \Phi(r_i, \sigma) \; \Phi(r_j, \sigma') \left[ \frac{1}{2} - \left| \frac{1}{2} - P(r_i - r_j) \right| \right]. \tag{39}$$

Calculating this quantity for the 2001-2005 seasons yields the results in Table 3.

| Season | Estimated Upset Frequency ($U$) |
|---|---|
| 2001 | 21.2% |
| 2002 | 21.3% |
| 2003 | 20.7% |
| 2004 | 21.6% |
| 2005[*] | 22.2% |

[*]Games through 11/19/05.

**Table 3.** Estimated Frequency of Upsets Obtained from the Model via the MSV and MTV for the Years 2001-2005.

---

[19] In the sense that the width of the posterior distribution described by the rating estimates alone (without consideration of their uncertainty) will be smaller than the width of the corresponding prior distribution. Thus, estimated rating differences will tend to underestimate the corresponding true rating differences.

We next wish to determine some reasonable upper and lower bounds on the upset frequency. Let us assume, for a moment, that the ratings produced by our model are good estimates of the mean performance levels of all of the teams. From the indicated values, we can determine the frequency of apparent upsets at various points in the season (i.e., Massey's "ranking violations" score). For our model, and typical of the ranking systems listed on Massey's ranking comparison page,[20] the ranking violations start out very low early in the season and climb steadily, presumably towards some asymptote that we would suggest is the "true [prior] frequency of upsets." Since the ranking violations from our model are fairly representative of the ranking systems covered by Massey, we will use it to depict this behavior. Using data from last complete season available as of this writing (2004), Figure 4 shows the frequency of ranking violations for our model as a function of the number of games considered (consecutively from the start of the season). The extrapolation shown is strictly notional – given the very limited data – and is meant only to demonstrate that an asymptote around 22% (per Table 3) is not unreasonable. In any case, end-of-season ranking violations (~16-19% for most ranking systems for Division I-A) provide an estimate of a *lower* bound on the "true" frequency of upsets.

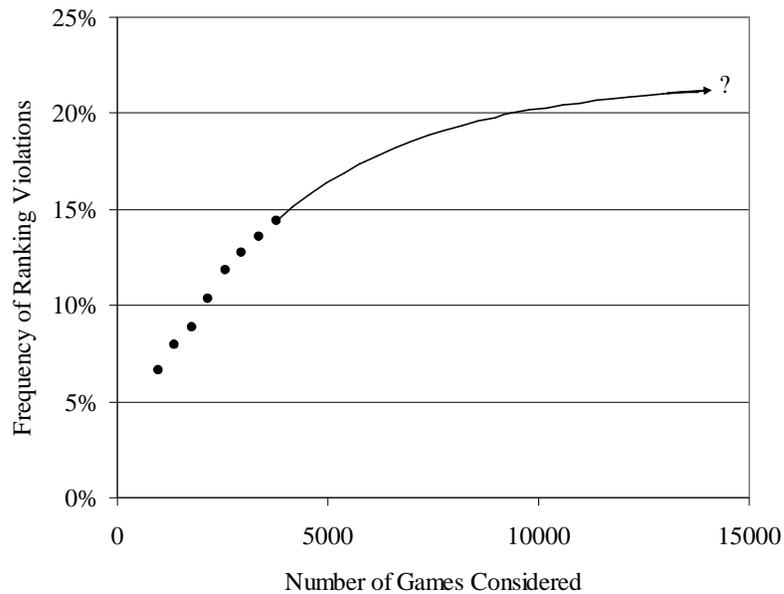

**Figure 4.** Frequency of Ranking Violations from the Model vs. Number of Games Considered for the 2004 Season (extrapolation is notional).

---

[20] Note that Massey's website includes statistics only for Divisions I-A (http://www.mratings.com/cf/compare.htm) and I-AA (http://www.mratings.com/cf/compare1aa.htm).

We now seek to estimate some *upper* bound on the frequency of upsets. Consider for a moment a hypothetical perfect rating system. Assuming that game-to-game performance variability is independent and therefore unpredictable to this otherwise perfect system, the upset frequency will define the frequency with which this perfect system's predictions[21] are in error. Put another way, the degree to which rating systems can predict game outcomes is limited by the frequency of upsets as driven by uncorrelated, and unpredictable, game-to-game variations in performance. Fair and Oster (2002) recently demonstrated that composite ranking systems (combining results from a number of individual ranking systems) are capable of predicting winners in Division I-A games on the order of 72% of the time. Thus, to the extent I-A games are representative of college football as a whole, we would estimate that upsets should occur no more than 28% of the time.[22]

Thus, these rough estimates of the upper (28%) and lower (16-19%) bounds on the upset frequency suggest that our estimates from the model (20-23%) are reasonable. This, in turn, suggests that the magnitudes of the MSV and MTV relative to the performance variability are also reasonable.

As we close our discussion on the analysis of the model, we wish to revisit our claim that a prior distribution centered on the mean opponent's rating is superior to one centered on zero. Again, if schedules were generated by randomly selecting opponents from all possible teams, then the two methods should be roughly indistinguishable. However, if opponents are selected from a more narrow range, then that should be reflected in the MSV when compared to the spread of ratings across all teams. This is, in fact, the case. Typically, the standard deviation of estimated ratings across all teams is on the order of 1.65 to 1.80 times the corresponding value of $\sigma'$. Note also, that the posterior distribution of ratings is estimated to be smaller than the prior by about 15% by the end of the season, thus amplifying the difference to about a factor of two. Some questions still remain concerning independents (i.e., teams without a conference affiliation), and further analysis will be required to determine how well they fit the model. Possible modifications might consist of evaluating separate values of the MSV and MTV for independents as a group.

---

[21] We are neglecting here the fact that some rating systems are explicitly predictive in nature while others are retrodictive. Consider that we are only seeking a rough estimate of an upper bound on the frequency of upsets.

[22] Massey's website includes both Division I-A (http://www.mratings.com/cf/compare.htm) and I-AA (http://www.mratings.com/cf/compare1aa.htm) ranking comparison pages. Typically, ranking violations are slightly lower on the I-AA page, suggesting that an upper bound of 28% on the upset frequency is still valid when lower divisions are included.

## 6. HOME FIELD ADVANTAGE

Up to this point, we have neglected the effects of home field advantage. While we are still examining approaches to modeling its effects, we will offer here what we consider to be our basic approach.[23] We begin by modifying equation (6) to include a home field advantage term,

$$\Delta_{ij}(r) = r - \tilde{r}_j + h_{ij} \tag{40}$$

where

$$h_{ij} = \begin{cases} +h, & \text{if } i \text{ is at home} \\ -h, & \text{if } j \text{ is at home} \\ 0, & \text{if at a neutral site,} \end{cases} \tag{41}$$

and $h$ is the mean home field advantage.[24]

To evaluate $h$, our basic approach is to use the following:

$$N_h = \sum_{\substack{all\ non-neutral \\ site\ played\ games}} P\left(\tilde{r}_{home} - \tilde{r}_{away} + h, \sqrt{\sigma_{home}^2 + \sigma_{away}^2 + \sigma_v^2}\right) \tag{42}$$

where $N_h$ is the total number of games where the home team wins. This expression is best solved for $h$ using Newton iteration. Typically, a single correction is made to $h$ for each iteration where the team ratings, rating uncertainties, MSV, and MTV are updated. Note in this equation that we are currently attempting to account for the uncertainty in the ratings of the two teams. For reference, the values of $h$ for 2001-2005 averaged 0.244 with a standard deviation of 0.016.[25]

Concerning the prior distribution, it is unclear whether or not it should be adjusted for home field advantage and equally unclear what form such an adjustment would assume. Excepting conference play, there is some tendency for well-matched teams to schedule pairs of games against each other (at alternating sites year-to-year) while games between mismatched teams are often isolated encounters at the site of the superior team (which is often better equipped to host the larger crowds associated with its fan base). In any case, any such correction is expected to have a minor effect on the final rating results.

---

[23] The equations to follow are tentative and is the focus of our current work.
[24] While some teams may be able to exploit home field advantage to a greater or lesser degree, this parameter exists only to "level the field" between teams that play more or fewer home games. Obviously, a predictive system might attempt to evaluate a home field advantage for each team.
[25] Using data from each season through the games considered in the final BCS ranking (typically games ending 3-8 December).

## 7. PUBLISHED RANKINGS

For purposes of our published rankings (http://www.atomicfootball.com), we utilize a slightly different ranking metric. In an ideal world, one might argue that every team would play every other team and the team with the highest winning percentage would simply be declared the winner. Since 700+ game schedules are obviously prohibitive, we produce instead an equivalent statistic as follows:

$$\tilde{\tilde{r}}_i = \left\langle P\left(\tilde{r}_i - \tilde{r}_j, \sqrt{\sigma_i^2 + \sigma_j^2 + \sigma_v^2}\right)\right\rangle_{all\ teams\ j \neq i} \tag{43}$$

where we have assumed a neutral site for all of these "hypothetical" games. Note that we have again attempted to account for the uncertainty in the team rankings (but have assumed here that the errors are uncorrelated). In this form, highly ranked teams will be penalized slightly for any factors that increase ranking uncertainty such as playing poorly matched opposition or inconsistent play – beating superior teams while losing to inferior ones. Otherwise, this ranking metric produces only slightly different results, occasionally flip-flopping two teams in the rankings. An interesting mathematical feature of this statistic is that it is independent of the value chosen for the relative performance variability. Note also that it takes the form of a winning percentage, falling in the range (0,1).

## 8. COMPARISON AND FINAL REMARKS

While we have downplayed the importance of comparisons with other rating systems, it may seem difficult to accept arguments as to the merits of any model without the benefit of at least a cursory comparison. For this purpose, we will include results from the model with and without the home field advantage adjustment given above. For our reference, we will use the BCS *computer* top ten rankings (*not* the composite BCS rankings that include the voting polls). The results shown in the tables below are coincident with the final official BCS computer rankings.[26]

---

[26] There is no BCS ranking, official or otherwise, after completion of the bowl games.

| 2001 Season |||| 2002 Season ||||
|---|---|---|---|---|---|---|---|
| BCS Computer Ranking | Team | Model Without Home Field | Model With Home Field | BCS Computer Ranking | Team | Model Without Home Field | Model With Home Field |
| 1 | Miami FL | 1 | 1 | 1 | Miami FL | 2 | 1 |
| 2 | Nebraska | 2 | 4 | 2 | Ohio State | 1 | 2 |
| 3 | Colorado | 4 | 3 | 3 | Georgia | 4 | 4 |
| 4 | Oregon | 3 | 2 | 4 | Southern Cal | 3 | 3 |
| 5 | Florida | 8 | 10 | 5 | Iowa | 7 | 8 |
| 6 | Tennessee | 5 | 5 | 6 | Oklahoma | 8 | 7 |
| 7 | Texas | 7 | 7 | 7 | Notre Dame | 6 | 6 |
| 8 | Stanford | 6 | 6 | 8 | Washington St | 5 | 5 |
| 9 | Oklahoma | 11 | 11 | 9 | Michigan | 9 | 10 |
| 10 | Illinois | 10 | 9 | 10 | Texas | 11 | 9 |

**Table 4.** Comparison Between the BCS Computer Rankings and Our Model With and Without Home Field Advantage for the 2001 and 2002 Seasons.

| 2003 Season |||| 2004 Season ||||
|---|---|---|---|---|---|---|---|
| BCS Computer Ranking | Team | Model Without Home Field | Model With Home Field | BCS Computer Ranking | Team | Model Without Home Field | Model With Home Field |
| 1 | Oklahoma | 1 | 1 | 1 | Oklahoma | 2 | 2 |
| 2 | LSU | 2 | 2 | 2 | Southern Cal | 1 | 1 |
| 3 | Southern Cal | 3 | 3 | 3 | Auburn | 3 | 3 |
| 4 | Michigan | 5 | 6 | 4 | Texas | 4 | 4 |
| 5 | Ohio St | 4 | 5 | 5 | Utah | 5 | 5 |
| 6 | Miami OH | 8 | 4 | 6 | California | 6 | 6 |
| 7 | Texas | 7 | 8 | 7T | Boise St | 7 | 7 |
| 8 | Florida St | 6 | 7 | 7T | Georgia | 9 | 11 |
| 9T | Miami FL | 10 | 9 | 9 | Virginia Tech | 10 | 9 |
| 9T | Tennessee | 11 | 11 | 10 | LSU | 11 | 12 |

**Table 5.** Comparison Between the BCS Computer Rankings and Our Model With and Without Home Field Advantage for the 2003 and 2004 Seasons.

| 2005 Season | | | |
|---|---|---|---|
| BCS Computer Ranking | Team | Model Without Home Field | Model With Home Field |
| 1 | Texas | 1 | 1 |
| 2 | Southern Cal | 2 | 2 |
| 3 | Penn State | 3 | 3 |
| 4 | Ohio State | 4 | 4 |
| 5 | Oregon | 6 | 6 |
| 6 | Virginia Tech | 5 | 5 |
| 7 | Miami FL | 8 | 7 |
| 8 | Georgia | 7 | 8 |
| 9 | West Virginia | 10 | 9 |
| 10T | LSU | 9 | 10 |
| 10T | Notre Dame | 14 | 13 |

**Table 6.** Comparison Between the BCS Computer Rankings and Our Model With and Without Home Field Advantage for the 2005 Season.

Overall, the matches are very good and appear to steadily improve from year to year.[27] We also found that our match with the computer average (as measured by the sum square of the ranking differences across all five years) is better than the median match among the current six BCS computer rankings. Since the current systems have the advantage of contributing to the average against which they are compared, recalculating the computer average with our model included as one of the systems (for all five years but including only the six BCS computer rankings current to 2005), our model has the best match by a comfortable margin (38.4 vs. 66.1) when home field advantage is disabled and the second best match by only a very small margin (62.9 vs. 62.6) when home field advantage is enabled.[28] Without going into more in-depth comparisons, we will simply state that to the extent that comparisons such as these have merit (at the very least as a "sanity check"), we believe the results of our model can be declared "very reasonable," as often seems to be the decree upon favorable comparisons of this kind.

In closing, we have described a completely self-contained statistical model for the ranking of college football teams using only win/loss information. Most importantly, we have achieved well-behaved results for undefeated and winless teams through the introduction of a Bayesian prior distribution for which the parameters describing it are calculated in a self-consistent manner. We have further demonstrated how those parameters can be validated against the actual game data. Finally, even though this model has no means by which it can be tuned, it shows remarkable agreement with the existing BCS computer rankings.

---

[27] Note that the 2001 season predated the BCS restriction to using only win/loss information.
[28] Not all of the current BCS computer rankings model home field advantage.

## 9. ADDENDUM (JUNE 2007)

In the Analysis section above, we speculated that examination of sets of three teams that each played the other two (what we shall call a "triplet") might be useful as a basis for a second metric by which the mean schedule variance (MSV) and mean team variance (MTV) can be established independently of the model. Since that time, we have become increasingly curious about the potential outcome of such an endeavor. Having succumbed to our curiosity, here is the product of that work.

First of all, we needed an appropriate metric. Guided by our original suggestion that we examine the relative frequency with which the three teams all go 1-1 (what we shall call a "split"), we derived the following:

$$Frequency\ of\ Splits = \frac{0.25}{1 + MSV + MTV} \qquad (44)$$

From this equation, we were able to determine an estimate for the frequency of splits from the MSV and MTV values derived by the ranking algorithm.

In evaluating the 2001 to 2005 seasons, we typically found about 6000 triplets within each season. Below is a summary of the results.

| | Frequency of Splits | | |
|---|---|---|---|
| Season | Estimated | Actual | Difference |
| 2001 | 9.24% | 9.09% | 0.15% |
| 2002 | 9.26% | 8.93% | 0.33% |
| 2003 | 8.78% | 8.94% | -0.16% |
| 2004 | 9.51% | 9.54% | -0.03% |
| 2005 | 9.84% | 9.44% | 0.40% |

**Table 7.** Frequency of Splits for all Division I-A, I-AA, II, III, and NAIA College Football Games for the Years 2001-2005 Derived from Actual Game Data and from Prior Distributions Obtained from the Model (Home Field Advantage Enabled).

We also estimated that a sample size of approximate 6000 triplets and a frequency of about 10% correspond to a precision of 0.39% (one sigma), consistent in magnitude with the RMS error of 0.25% in our relatively small sample.

We previously had indicated concerns that the matchups in these triplets might not be representative of the population. To further investigate this, we calculated the RMS rating difference for all games and compared this to the RMS rating difference for games involved in the triplets. For the 2005 season, the RMS rating difference for all games was 1.067 while the RMS rating difference for triplet games was 1.036. We estimate that differences on this order would cause the actual split frequency to be increased by about 0.3 to 0.4%.

Furthermore, we have only just begun to explore the influence of home field advantage. The effects are complicated by the fact that home field is not necessarily random, but is instead correlated with the relative strength of the teams (traditionally stronger teams, at least in nonconference games, are somewhat more likely to play at home). Assuming home field is random, we would estimate that it would reduce the actual split frequency by about 0.2 to 0.3%. Any correlation would lessen this.

In any case, the two perturbations mentioned above are relatively small and tend to negate each other. Neglecting them for now, our results are still very encouraging given that the differences between predicted and actual values are small, consistent with the precision estimates, and apparently free of any significant biases.

## 10. RELATED WORKS